\documentclass[submission, Proceedings]{SciPost}

\hypersetup{
    colorlinks,
    linkcolor={red!50!black},
    citecolor={blue!50!black},
    urlcolor={blue!80!black}
}

\usepackage[bitstream-charter]{mathdesign}
\DeclareSymbolFont{usualmathcal}{OMS}{cmsy}{m}{n}
\DeclareSymbolFontAlphabet{\mathcal}{usualmathcal}

\begin{document}

\begin{center}{\Large \textbf{
The Giant Radio Array for Neutrino Detection (GRAND) Project\\
}}\end{center}

\begin{center}
B. L. Lago\textsuperscript{1$\star$} for the GRAND collaboration
\end{center}

\begin{center}
{\bf $^1$} Centro Federal de Educa\c{c}\~ao Tecnol\'ogica Celso Suckow da Fonseca, Nova Friburgo, Brazil
\\
$\star$ bruno.lago@cefet-rj.br
\end{center}

\begin{center}
\today
\end{center}


\definecolor{palegray}{gray}{0.95}
\begin{center}
\colorbox{palegray}{
  \begin{tabular}{rr}
  \begin{minipage}{0.1\textwidth}
    \includegraphics[width=30mm]{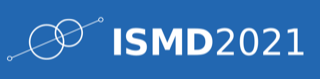}
  \end{minipage}
  &
  \begin{minipage}{0.75\textwidth}
    \begin{center}
    {\it 50th International Symposium on Multiparticle Dynamics}\\ {\it (ISMD2021)}\\
    {\it 12-16 July 2021} \\
    \doi{10.21468/SciPostPhysProc.?}\\
    \end{center}
  \end{minipage}
\end{tabular}
}
\end{center}

\section*{Abstract}
{\bf
GRAND is designed to detect ultra-high-energy cosmic particles -- specially neutrinos, cosmic rays and gamma rays using radio antennas. With $\sim$20 mountainous sites around the world it will cover a total area of 200,000 km$^{2}$. The planned sensitivity of 10$^{-10}$ GeV cm$^{-2}$ s$^{-1}$ sr$^{-1}$ above $5\times10^{17}$ eV will likely ensure the detection of cosmogenic neutrinos predicted by most common scenarios enabling neutrino astronomy. Furthermore, PeV--EeV neutrinos can test particle interactions at energies above those achieved in accelerators. The pathfinder stage GRANDProto300 is planned to start taking data in 2021. We present the current overall status of the project with emphasis on the neutrino physics.
}


\section{Introduction}
\label{sec:intro}

Ultra-high-energy cosmic rays (UHECRs) are extraterrestrial charged particles that reach Earth with energies above 10$^{18}$ eV (1 EeV). Their origin is likely to be extragalactic \cite{Aab_2017} although no sources have been identified so far. The search for astrophysical objects that could accelerate particles to such high energies is usually based on arrival direction studies.

The detection of the UHECRs arriving at Earth presents numerous challenges to be overcome. In particular, we do not know the mass composition on an event by event basis. Moreover, since the particles are charged, they are deflected both in the galactic and extragalactic magnetic fields, so that they do not point back to their sources. Furthermore, the cosmic ray flux is tiny for higher energies resulting in few events above 40 EeV from sources beyond 100 Mpc. See \cite{Alves_Batista_2019, Schroeder2019High} for a review on the UHECRs state of the art.

Another possible avenue to understand cosmic accelerators is to look for EeV gamma rays and neutrinos produced by UHECRs in their propagation stages or at the sources. Although the expected flux of these cosmogenic particles is below the detection limits achieved thus far, they are guaranteed to exist. Since $\gamma$ and $\nu$ are neutral, their trajectories are not deflected by the magnetic fields and they point back at their sources. There is a limit on the source distance for gamma rays since they do not reach Earth from beyond 10 Mpc due to interaction with CMB. On the other hand, the universe is transparent to UHE cosmogenic neutrinos.

Given that cosmogenic gamma rays and neutrinos can provide important information on the composition, spectrum and redshift evolution of the sources of UHECRs \cite{Batista_2019,Sarkar_2011}, it is worth to aim for a detector sensitive to these messengers.

\section{Science Case}
\label{sec:science}

The Giant Radio Array for Neutrino Detection (GRAND) project has a wide and rich science case \cite{GRAND_2020}. In the energy range from $10^{7.5}$ GeV to $10^9$ GeV there is a transition, from galactic to extragalactic, on the origin of UHECRs. The early stage of GRAND will detect cosmic rays in this energy range providing high statistics. Furthermore, with an aperture of 107,000 km$^2$ ($\sim$ 20 times the current aperture of Auger) GRAND will reach 20-80 times the exposure of Auger featuring full sky coverage on the latest stages. GRAND will help shed light into topics such as: muon discrepancy, UHECR mass composition and p-air cross section.

Neutrinos play an important role on the detection principle of GRAND. Hence, neutrino astronomy, neutrino cosmogenic flux, neutrino cross-section measurements, spectral and/or angular distribution distortions and flavour ratios are among the topics addressed by GRAND.

The same process involved in the production of cosmogenic neutrinos yields UHE gamma rays. The  predicted sensitivity for such gamma rays in GRAND will be competitive since the early stages of development.

GRAND will be able to detect Fast Radio Bursts \cite{Masui_2015} and Giant Radio Pulses \cite{Soglasnov_2006} by incoherently adding the signals from individual antennas. The temperature of the sky will be measured with mK precision, as a function of frequency. Hence, GRAND will reconstruct the global Epoch of Reionization signature and identify the absorption feature due to reionization below 100 MHz following the approach described in \cite{Bowman_2018}.

\section{Detector: principle and design}
\label{sec:detector}

The detection principle is based on the radio signal emitted by the Extensive Air Showers (EAS) produced by the UHECRs in the atmosphere. The radio emission occurs due to the interaction of mainly electrons and positrons with the geomagnetic field and the excess of electrons established during the development of the EAS. This radiation can be detected far from the shower since it is not attenuated much by the atmosphere. For more details see \cite{Huege_2016}.

Cosmic rays with incoming trajectories closer to the horizon produce a larger footprint on the ground and GRAND is well designed to detect such events. Cosmic rays will be detected by reaching the Earth and inducing an EAS. Cosmogenic neutrinos produced by them in their propagation through the Universe can also produce EAS close to the horizon. There are two possibilities for the neutrino detection: ($i$) a tau neutrino interacting underground or inside a mountain produces a tau particle that escapes and decays in the atmosphere above the GRAND site. ($ii$) A similar process may occur (less probable than the previous one) for very horizontal events since the atmosphere traversed in this case is roughly equivalent to 100 m of rock.

The GRAND design is modular and aims to cover an area of 200,000 km$^2$ with 20 sites spread worldwide. Each site is panned to have 10,000 antennas and to be located on radio quiet mountainous regions. The irregular landscape increases the probability of tau neutrinos interacting inside mountains and the produced tau lepton decaying in the atmosphere also improves the detection rate when compared to a flat landscape. With sites all around the globe GRAND will have full sky coverage.

The antennas are relatively cheap and will operate on the 50 MHz to 200 MHz frequency range. The bow tie design was chosen to enable flat response in both azimuthal angle and frequency. Since earth-skimming events are expected to have zenith angles close to 90$^\circ$, the antennas will be placed at a height of 5 m in order to avoid interference due to reflection of the radio signal on the ground.

\section{Ultra-high-energy messengers}
\label{sec:messengers}

\subsection{Neutrinos}

\begin{figure}[h]
  \begin{minipage}[b]{0.48\textwidth}
    \centering
    \includegraphics[width=\linewidth]{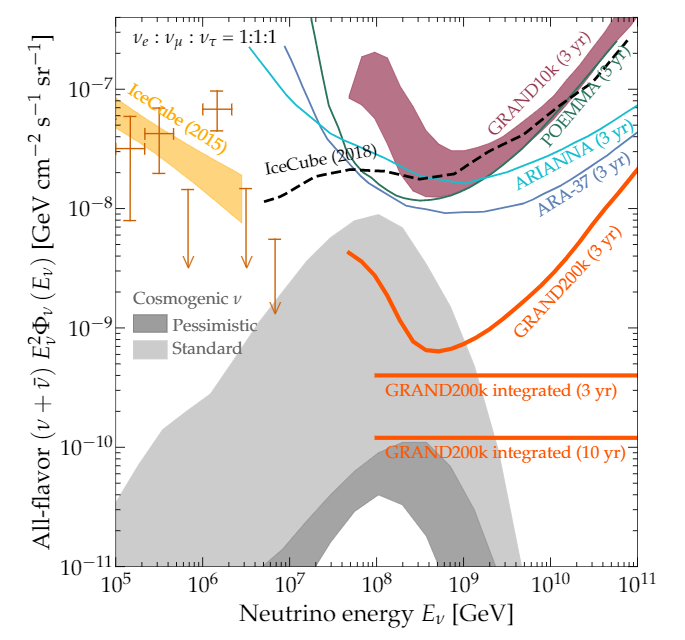}
   \end{minipage} 
 \hfill
 \begin{minipage}[b]{0.48\textwidth}
  \centering
  \includegraphics[width=\linewidth]{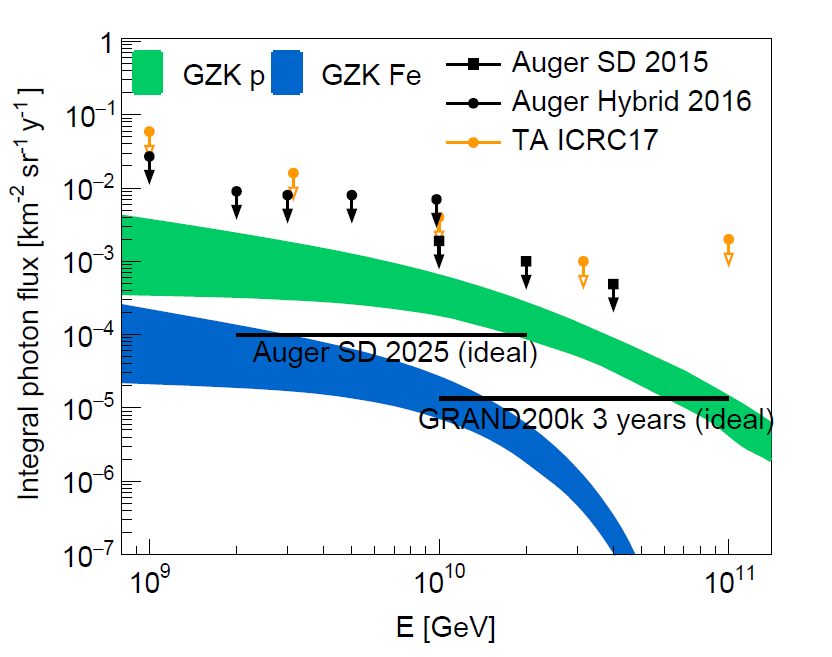}
 \end{minipage} 
 \caption{\emph{Left}: Predicted cosmogenic neutrino flux, compared to experimental upper limits and sensitivities of several existing and projected experiments. The gray-shaded regions were obtained through the combined fit of energy spectrum and mass-composition using Auger data and UHECR simulations \cite{Batista_2019}. \emph{Right}: Estimated upper limits on UHE photon sensitivity for GRAND after 3 years of operation. For reference, the predicted cosmogenic UHE photon flux for pure-proton and pure-iron scenarios, as estimated in \cite{Sarkar_2011} was also shown. For comparison, we include the existing upper limits from Auger and the Telescope Array (TA) and the projected reach of Auger by 2025.}
 \label{fig:nu_and_gamma}
\end{figure}

The interaction of UHECRs with either the Cosmic Microwave Background (CMB) or the Extragalactic Background Light results in a diffuse flux of cosmogenic neutrinos. The predictions for the flux of such neutrinos suffers from a large spread due to: Properties of the UHECRs, distribution and evolution with redshift of sources, neutrino emissivity of the source, UHECR injection spectrum and mass composition of the injection at the source. Nevertheless, in its latest stage GRAND will achieve enough sensitivity ($E^2 {\rm d}N/{\rm d}E_\nu\sim$10$^{-10}$ GeV cm$^{-2}$ s$^{-1}$ sr$^{-1}$) to detect these neutrinos, even in the pessimistic scenarios (figure \ref{fig:nu_and_gamma}). UHECRs interacting with photons and hadrons inside the sources can produce neutrinos. After 3 years of operation, GRAND could discover the first sources of UHE neutrinos at a significance of 5$\sigma$ \cite{Fang_2016}.

Given that several new physics models have effects that scale like $\kappa_n E_\nu^n L$, where $E_\nu$ is the neutrino energy and $L$ is the source-detector baseline, GRAND is likely to probe $\kappa_n$ down to 4$\times$10$^{-50}$ EeV$^{1-n}$ for $E_\nu=$1 EeV and $L=$ 1 Gpc. This sensitivity is orders of magnitude greater than the ones obtained from atmospheric and solar neutrinos \cite{Abbasi_2010, Abe_2015}.

GRAND will measure the angular distribution of the tau neutrino flux around the Earth with a sub-degree angular resolution. Assuming the incoming neutrinos follow an isotropic distribution, it is possible to extract the tau neutrino-nucleon cross section \cite{Denton_2020}. 
Another observable sensitive to new physics with neutrinos is the spectral shape. The search for features, peaks and slope changes on the neutrino energy spectrum benefits from GRAND’s high statistics and energy resolution. For a brief review on the neutrino topic see \cite{Bustamante_2019}.

\subsection{Gamma rays}

Although not detected so far, UHE gamma rays are also guaranteed to exist. Given the predicted sensitivity (see figure \ref{fig:nu_and_gamma}) of GRAND to UHE gamma rays the task goals are:
\begin{itemize}
\item{Measure the flux of cosmogenic gamma rays above 10$^{10}$ GeV or strongly constrain it.}
\item{Detect nearby sources of UHE gamma rays up to to 10Mpc since gamma rays point back to their astrophysical sources.}
\item{The detection of UHE gamma rays would probe the little known diffuse cosmic radio background (CRB). GRAND could be the first experiment to put such indirect constraints on the local (up to 10Mpc) CRB with its full efficiency for photon detection in the energy range from 10$^{10}$ to 10$^{11}$ GeV.}
\end{itemize}

\subsection{Cosmic rays}

GRAND was designed to provide a precise measurement of the shape of the cosmic ray energy spectrum cutoff around 4$\times 10^{10}$ GeV. This piece of information will enable the distinction between scenarios where the sources run out of power and those where GZK processes are dominant \cite{Kotera_2011}. In the latest stages GRAND will be fully sensitive to cosmic rays with energies greater than $10^{10}$ GeV and zenith angles $65^\circ \leq \theta \leq 85^\circ$ resulting in 32,000 events with energies above $10^{10.5}$ GeV after 5 years of operation.

The number of muons detected from EAS produced by cosmic rays with energies above 10$^{9.5}$ GeV is higher than the one predicted by air-shower simulations \cite{Aab_2015}. GRAND will combine independent measurements of the electromagnetic component — which depends
less strongly on hadronic interaction models — and the muon component. That approach could disentangle the differences in shower development due to different choices of hadronic interaction models and mass compositions, thus alleviating one of the main sources of uncertainty when inferring the mass composition of UHECRs in ground arrays \cite{Aab_2016, Aab_2017b}.

GRAND will feature a full sky coverage with both the Telescope Array hotspot \cite{Abbasi_2014} and the Pierre Auger dipole \cite{Aab_2017} within its field of view. Full sky coverage allows for a better reconstruction of anisotropies than partial sky coverage even for the same total event rate \cite{Denton_2015}. Furthermore, the high statistics will enable a thorough study of cosmic ray arrival directions.

Given the high number of events detected by GRAND, even in a scenario with low fraction of incoming protons, it will be possible to measure the proton-air cross section up to $\sqrt{s}\sim2\times 10^5$ GeV extending the results from \cite{Abreu_2012, Abbasi_2020}.

\subsection{Multimessenger}
By the time GRAND reaches its later stages, other experiments will be further developed and several messengers may be available: neutrinos, cosmic rays, photons, and gravitational waves. Future experiments will observe mergers from cosmological distances and have a wide coverage of the electromagnetic spectrum.

UHE neutrinos from transient point sources detected by GRAND, combined with electromagnetic observations performed by other experiments, will benefit from GRAND’s sub-degree angular resolution and timing, making it possible to pinpoint neutrino sources from galaxies within the field of view.

The UHE gamma rays from transient point sources produce synchrotron radiation through their interaction with the Large Scale Structure around the source. It is possible to perform joint analysis since the former can be detected by GRAND and the latter by other experiments. Synchrotron emission with energies greater than 0.1 TeV produced in magnetized structured regions where the sources are embedded \cite{Murase_2012} may be detected by Cherenkov Telescope Array - CTA \cite{CTA_2018}. The  Large High Altitude Air Shower Observatory (LHAASO) reported events with energies from  100 TeV up to 1.4 PeV \cite{Cao_2021}. Another possible joint analysis combines gamma rays from short Gamma Ray Bursts detected by GRAND and gravitational waves detected by the Laser Interferometer Gravitational-Wave Observatories - LIGO, VIRGO and KAGRA \cite{Abbott_2020} (emitted hours or days earlier).

Due to its unprecedented UHE neutrino sensitivity, GRAND will be a crucial triggering and follow up partner in multimessenger programs.

\section{GRANDProto300}
\label{sec:GP300}

The next stage of GRAND is the 300-antenna pathfinder GRANDProto300 \cite{ZhangYi_2021}. It will cover an area of 200 km$^2$ with radio antennas, 100 of which have been produced and are ready to be deployed. It aims to validate GRAND as a standalone radio detection array and realize the self-trigger techniques. It will also be a test bench to improve angular, energy and mass composition reconstructions.

The focus of this stage is the detection of very inclined events with energies from 30 PeV to 1 EeV in order to study the galactic to extragalactic transition on the origin of cosmic rays. It has potential sensitivity for radio transients such as Giant Radio Pulses and Fast Radio Bursts.

In the current GRAND schedule GRANDOProto300 is planned to begin its operations this year.

\section{Conclusion}

GRAND was designed to be the largest UHE particle observatory in the world with improved statistics and full sky coverage. A modular design using relatively inexpensive antennas make it possible to have sites all around the world and cover a larger area. It will be sensitive to cosmic rays, cosmogenic neutrinos and cosmogenic gamma rays making it possible to perform multimessenger studies, specially when combined with other experiments. In the near future GRANDProto300 will begin its operation studying the galactic to extragalactic transition and serving as a pathfinder for the later stages.

\section*{Acknowledgements}
I thank Prof. Jaime Álvarez-Muñiz, Dr. Rafael. A. Batista, Prof. João R. T. de Mello Neto and Dr. Peter B. Denton for useful suggestions and comments on this paper and Prof. Mauricio Bustamante for discussions on the topic.


\bibliography{References.bib}

\nolinenumbers

\end{document}